\documentclass[showpacs,letterpaper,12pt]{revtex4-1}
\usepackage{times}
\usepackage{amsmath,bm,amsfonts}
\usepackage{dcolumn}
\usepackage{graphicx}
\usepackage{mathrsfs}
\usepackage{latexsym}

\begin{document}

\title{Theory of Superconductivity in Strongly Correlated Electron Systems}
\author{Chyh-Hong Chern}
\email{chchern@ntu.edu.tw, chern@alumni.stanford.edu}
\affiliation{Department of Physics, 
National Taiwan University, Taipei 10617, Taiwan}
\date{14 August 2018}

\begin{abstract}
In the correlated electron system with the pseudogap, there are full-gapped domains and Fermi-arced domains coexisting.  Those domains are created by the quantum-fluctuated antiferromagnetic correlation that generate the short-ranged attractive potential to produce the Fermi arcs and the superconductivity.  In the full-gapped domains, $s$-wave or ($d_{x^2-y^2}\pm id_{xy}$)-wave symmetry of the electron pairs is favored.  In the Fermi-arced domains, only $d_{x^2-y^2}$-wave symmetry of pairs is stable.  Superconductivity of different pairing symmetry coexists in different domains, as well.  Different from the Cooper pairs, the correlated electrons pair up in the \emph{real} space with an energy gap.  Gapless states, on the contrary, hinder the development of superconductivity.\end{abstract}

\pacs{74.20.-z, 74.20.Mn, 74.20.Rp}

\maketitle

\section{Introduction}

Electronic structure and topology in materials dictate their classification.  For example, a system with (without) a Fermi surface is a metal (an insulator).  The systems with the Fermi surface have gapless low-lying excitations, that are usually not stable against perturbations, leading to various states of matter through phase transitions.  Superconductivity arises in metals when there is Cooper-pair instability.  Interacting with phonons~\cite{bardeen1957a, bardeen1957b}, electrons acquire the attractive force to form the bosonic pairs.  The pairs, in particular, are robust in any directions.  Namely, the two-particle pairing wavefunction has zero angular momentum, resulting in an isotropic superconducting gap. The presence of the Fermi surface is the necessary condition for Cooper instability.  In other words, the gapless states are the nurturing cradle for superconductivity

That elegant scenario was challenged by the discovery of the high transition temperature superconductivity in copper oxides in 1986~\cite{bednorz1986}.  The new superconductivity appears in families of compounds that mobile charge carriers need to be doped.  Taking La$_{\text{2-}x}$Sr$_{x}$CuO$_{4}$ as an example, at $x$=0, the compound is an antiferromagnetic insulator with an antiferromagnetic transition temperature up to near 300 K~\cite{vaknin1987}.  Increment of $x$ introduces the mobile holes as well as the gapless states in the $(\pm\frac{\pi}{2},\pm\frac{\pi}{2})$ directions~\cite{zhen2003}.  They form the arc segments, dubbed as Fermi arcs.  A closed Fermi surface can be completed at the doping level around 19\%~\cite{taillefer2007}.  While the Fermi arcs grow in the $(\pm\frac{\pi}{2},\pm\frac{\pi}{2})$ directions, the directions along $(\pm\pi,0)$ and $(0,\pm\pi)$ remain gapped.  

The coexistence of the gapless and gapped states blurs the boundary of metals and insulators.  So far, the origin of the Fermi arc formation still stays at the numerical evidence~\cite{imada2009}.  The clear physical mechanism is lacking.  Meanwhile, in those very strange electronic structures, superconductivity arises.  Two properties remain true.  Namely, electrons form pairs, and most of their properties are well described by BCS wavefunction of $d$-wave pairing symmetry.  However, the $d$-wave symmetry that electrons choose to pair is $d_{x^2-y^2}$, which is very odd.  The $d_{x^2-y^2}$ state has the maximum amplitude in the $(\pm\pi,0)$ and $(0,\pm\pi)$ directions and are zero in the $(\pm\frac{\pi}{2},\pm\frac{\pi}{2})$ directions.  That choice violates our general understanding.  Based on Cooper, electrons pair up in the momentum space where there are gapless states. Then, $d_{xy}$ should be favored.  The special choice of $d_{x^2-y^2}$ symmetry implies that the electron pairs in cuprates are \emph{not} Cooper pairs.

In this paper, we provide the physical mechanism of Fermi arc formation and the pairing mechanism that explains why $d_{x^2-y^2}$ symmetry is favored.  We will show that the correlated electrons are in general not homogeneous.  Namely, domains with Fermi arcs and the ones without can coexist.  Likewise, $d_{x^2-y^2}$, $d_{x^2-y^2}\pm id_{xy}$, and $s$ symmetries of pairing can coexist in different domains \emph{in the same system}, as well.  The physics measured depends on the experimental techniques.  For examples, in the angle-resolved photoemission experiments, light often shines on the area of multiple domains.  In this case, the photo-electrons from the gapless states always come out, if present.  This technique often claims the $d_{x^2-y^2}$ symmetry.  For the scanning tunneling microscopy experiments and other techniques of local probe, they can claim all cases, depending on which domain they measure~\cite{imada2013}.  Different from Cooper pairs, the superconductivity in the strongly correlated electron system can arise \emph{without} a Fermi surface, {\it i.e.} fully gapped, because electrons pair up in the real space, not in the momentum space.

\section{Fermi arc formation}
The renowned electronic structure in cuprates, the Fermi arcs, is the combinations of the two mechanisms: pseudogap formation and quantum fluctuations of the correlation degrees of freedom.  The first one is due to the interaction of electrons through exchanging massive gauge bosons of \emph{imaginary} wave vectors.  The second one is the interaction between electrons and the gauge field of \emph{real} wave vectors.  In comparison to the electrodynamics, the first one is similar to the Coulomb interaction of exchanging virtual photons, and the second one is the additional interaction with electromagnetic waves.  The first mechanism results in the repulsive interaction between electrons and opens a gap.  The second mechanism creates an attractive potential that cancels the repulsive interaction and closes the gap in some circumstances.  Based on the form factors of the attractive potential, it closes the gap in an anisotropic manner, leading to the arc segments in the momentum space.  In this section, we discuss the effect of the second mechanism.  We will show that the attractive potentials have two types: $s$-type and $d$-type.  The $s$-type remains the pseudogap to be a full gap, and the $d$-type closes the gap in the nodal directions. 

Let us begin with a brief introduction of the pseudogap formation.  The antiferromagnetic fluctuation in cuprates leads to divergent phenomena that seemingly do not have any relations~\cite{mendels1989, julien2011, keimer2012, kapitulnik2008}.  Assuming its $XY$ nature at finite doping, it is disordered in high temperature and acquires a quasi-long-ranged order through the Kosterlitz-Thouless transition~\cite{chern2014}.  In addition, the overlap of the electron spin wavefunction introduces the massless gauge field strongly coupling with the antiferromagnetic fluctuation, that provides the mass of the gauge field and becomes the longitudinal mode of the gauge field~\cite{wen1989, chern2014}.  Electrons weakly coupling with the massive gauge field open a gap in the excitation spectrum, which is the origin of the pseudogap formation.  

In the pseudogap phase, the quantum fluctuation of the anfiferromagnetic fluctuation can occur~\cite{lee2018}.  It generates a propagating $\vec{E}$ field.  In the domain that the fluctuation stably exists and assuming the $x$ direction to be the longitudinal direction, the propagating $\vec{E}$ field forms a standing wave given by
\begin{eqnarray}
&&\vec{E}=-\frac{A_0M^2_0}{k_L}\sin(k_Lx)\cos(\omega_L t)\hat{x}, \label{e-field}
\end{eqnarray}
where $A_0$ is the strength of the quantum fluctuation, $M_0$ is the mass of the gauge field, and $\omega^2_L=k_L^2+M_0^2$.  The electric field in Eq.~(\ref{e-field}) drives the electrons and form the charge density modulations~\cite{lee2018}.  Taking the onset temperature of the density modulation $T_{\text{CDW}}$, $A_0$ can be estimated as $A_0=\sqrt{\frac{16\pi^2k_BT_{\text{CDW}}}{4\pi^2+1}}$.

\begin{figure}[htb]
\includegraphics[width=0.48\textwidth]{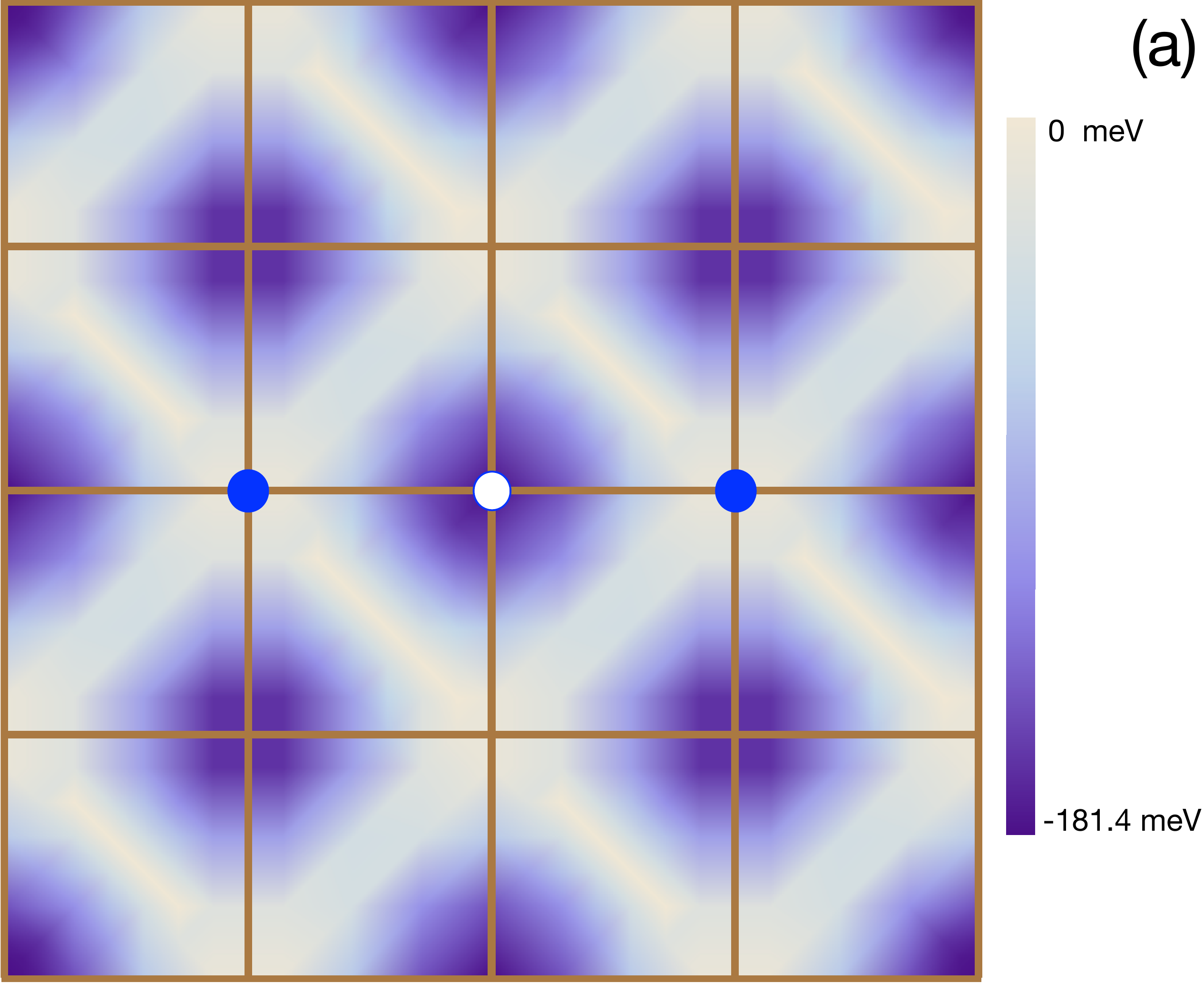}
\includegraphics[width=0.48\textwidth]{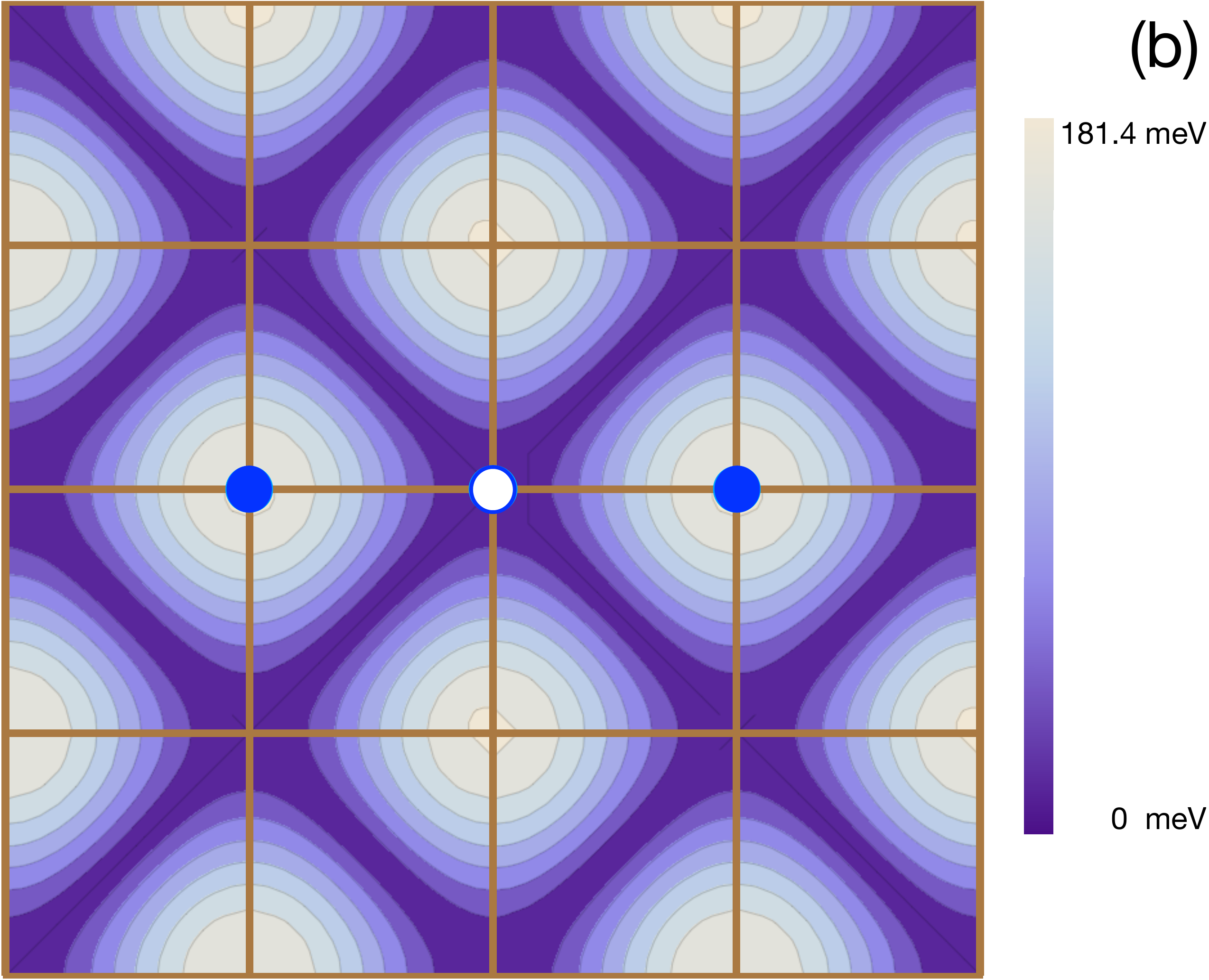}
\caption{(Color online) The effective potential of the antiferromagnetic fluctuations.  Brown lines indicate the lattice of copper atoms. The blue empty circle indicates the hole site.  The blue solid circles indicate the electron sites.  (a) $\bar{V}_1(x,y)$. (b) $\bar{V}_2(x,y)$.}\label{potential}
\end{figure}

In general, the antiferromagnetic fluctuation can be equally excited in the $\hat{y}$ direction of the square lattice.  In order to match the locations of the nodes in the $x$ direction, there are only two possibilities.
\begin{eqnarray}
&&\vec{E}_1(x,y)=-\frac{A_0M^2_0}{k_L}\cos(\omega_L t)[\sin(k_Lx)\hat{x}+\sin(k_Ly)\hat{y}], \nonumber\\
&&\vec{E}_2 (x,y)=-\frac{A_0M^2_0}{k_L}\cos(\omega_L t)[\sin(k_Lx)\hat{x}-\sin(k_Ly)\hat{y}],
\end{eqnarray}
that are equivalent to the potentials
\begin{eqnarray}
&&V_1(x,y)= -V_0\cos(\omega_L t)[\cos(k_Lx)+\cos(k_Ly)], \nonumber \\
&&V_2 (x,y)= -V_0\cos(\omega_L t)[\cos(k_Lx)-\cos(k_Ly)],
\end{eqnarray}
where $V_0=\sqrt{\frac{6m^*\Delta k_BT_{\text{CDW}}}{(4\pi^2+1)k_L}}$, $m^*$ is the effective electron mass, and $\Delta$ is the pseudogap magnitude.  Taking $\Delta\sim 40$ meV~\cite{zhen2014} and $T_{\text{CDW}}\sim$ 100 K, $k_L$ for the 4 lattice spacings modulation, $V_0$ = 0.363 eV.  The classical dynamics of the electrons in the presence of the quantum antiferromagnetic fluctuation is already given in Ref.~\cite{lee2018} in details.  In addition, the rapidly oscillating $\vec{E}$ field drives the electrons to move around the nodes, which effectively create the attractive potentials~\cite{lee2018}
\begin{eqnarray}
&&\bar{V}_1 (x,y)= -V_0|\cos(k_Lx)+\cos(k_Ly)|, \nonumber \\
&&\bar{V}_2 (x,y)= V_0|\cos(k_Lx)-\cos(k_Ly)|,
\end{eqnarray}
as shown in Fig.~\ref{potential}.  The potential minimum indicate the location of the nodes.  In $\bar{V}_1 (x,y)$, the nodes are the local minimum.  In $\bar{V}_2 (x,y)$, the nodes are surrounded by the lines of minimum.

Now, let us consider the doping where one of the electrons at the nodes is taken away as shown in Fig.~(\ref{potential}).  The electrons at the anti-nodes, shown as the solid blue circles in Fig.~(\ref{potential}), then move toward each other and oscillate around the nodes~\cite{lee2018}.  The attractive force in the potential is larger than the repulsive force that generates the pseudogap, which can be estimated as the following.   The attractive force is roughly estimated from the depth of the effective potential 0.363 eV with a length scale $\frac{1}{4M_0}$, and the repulsive force can be done from the 40 meV pseudogap with a length scale $\frac{1}{M_0}$.   Therefore, the attractive force cancels the repulsive force and closes the pseudogap dynamically.  In the case of potential $\bar{V}_2$, two electrons can scatter off in any directions, as they collide.  If they scatter off to the anti-nodes, they repeat the motion again.  If they scatter off along the lines of minimum, the nodal quasiparticles appear, since there is no confining potential.  The direction of the lines of the minimum is nothing but that of $(\pm\frac{\pi}{2},\pm\frac{\pi}{2})$, that provides the reason why Fermi arcs form in those directions.  On the other hand, in the case of the potential $\bar{V}_1$, electrons always scatter off back to the same position and repeat the oscillations.  There is no gapless excitation.   Finally, we emphasize that the current scheme is different from the excitation process between bands where there is no force to cancel the repulsive many-body interaction.  In conclusion, the prevailing antiferromagnetic fluctuation and its form factor dictate the Fermi arc formation in the $(\pm\frac{\pi}{2},\pm\frac{\pi}{2})$ directions.

Both $\bar{V}_1$ and $\bar{V}_2$ are doping dependent.  $V_0$ is the function of pseudogap magnitude $\Delta$ and the onset temperature of the density modulation $T_{\text{CDW}}$.  As the doping increases, $\Delta$ decreases.  The depth of the potential decreases with the doping, and electrons can scatter off in wider directions, leading to the growth of the Fermi arcs.  This mechanism implies that the potentials can be still finite when the Fermi arcs complete Fermi surfaces in both $\bar{V}_1$ and $\bar{V}_2$ cases.  Namely, the completion of the Fermi surfaces does not implies that $\Delta$ is zero~\cite{zhen2014}.

The existence of the potential $\bar{V}_1$ and $\bar{V}_2$ should be of equal probability.  Namely, they can coexist in different domains in the same system.  The pseudogap structures are inhomogeneous in the real space and can be classified into two categories.  (1) full pseudogap (from $\bar{V}_1$), and (2) Fermi-arced pseudogap (from $\bar{V}_2$).  Although the domain sizes can not be estimated, we believe that physics measured highly depends on the experimental tools.  For example, the photo-emission experiments and the transport measurements are sensitive to the gapless excitations.  As their probes cover multi-domains, the physics from $\bar{V}_2$ dominates the signals.  For the tools of local probes, especially STM, physics of both pseudogap structures should be faithfully claimed.  

\begin{figure}[htb]
\includegraphics[width=0.48\textwidth]{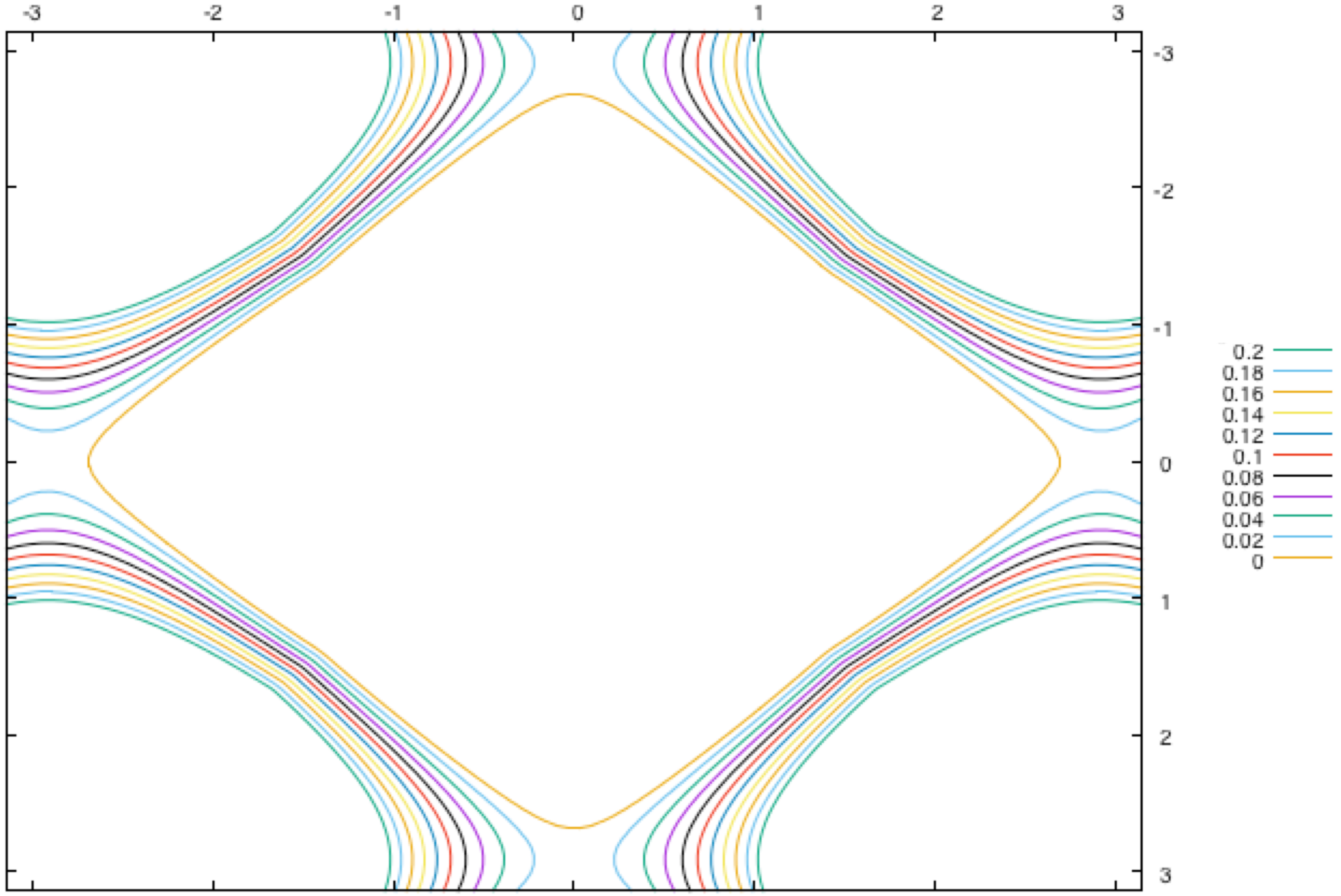}
\caption{(Color online) The equipotential contours of the band in Eq.~(\ref{band}).  The points of $(\pm\frac{\pi}{2},\pm\frac{\pi}{2})$ locate at 0.128 of the equipotential.  }\label{contour}
\end{figure}

Next, let us construct the band structure observed in the photo-emission experiments.  With no proof, we find that the bands can be phenomenologically described by $\mathscr{E}(k_x, k_y) = \mathscr{E}_1+\frac{1}{2}(\mathscr{E}_2+\mathscr{E}_3)$, where
\begin{eqnarray}
&&\mathscr{E}_1 = -2t(\cos k_x + \cos k_y), \nonumber \\
&&\mathscr{E}_2 = V_0|\cos \frac{k_x}{2} + \cos \frac{k_y}{2}|, \nonumber \\
&&\mathscr{E}_3 = -V_0|\cos \frac{k_x}{2} - \cos \frac{k_y}{2}|, \label{band}
\end{eqnarray}
$t$ = 0.25 eV, and $V_0$ = 0.363 eV.  Eq.~(\ref{band}) satisfies the dispersion relation in the Fig.(1) of Ref.\cite{zhen2014}.  The first term is from the tight-binding model in the square lattice.  The second and third terms are from the antiferromagnetic fluctuation, where the periodicity is double in both $x$ and $y$ directions.  It is not surprising that $\bar{V}_1$ and $\bar{V}_2$ contribute to the band structure.  The energy contours of equipotentials are given in Fig.~(\ref{contour}).  The introduction of the $\mathscr{E}_2$ and $\mathscr{E}_3$, namely the quantum fluctuation of the antiferromagnetic fluctuation, changes the electron-like band to the hole-like band.  Further doping can change back to the electron-like band again.  The change of the topology of the band structure can be observed in the measurements of the Hall coefficient~\cite{taillefer2011}.


\section{Superconductivity}

Superconductivity arises at finite doping where the Fermi-arcs appear.  It sounds reasonable for the superconductivity in the BCS theory.  However, superconductivity in correlated electron systems is not the case.  Despite the reason given in the introduction, the superconductivity diminishes after Fermi surfaces complete.  Namely, the superconductivity in cuprates does not favor gapless states.  Furthermore, the electron pairs can be $d$-wave symmetry.  Where does the angular momentum come from?  

BCS theory conveys two concepts: (1)  Cooper's instability due to the interaction with phonons (2) BCS wavefunction and its reduced Hamiltonian.  Although the first one does not apply to the correlated electron systems, the second one remains true.  Therefore, the question of the superconductivity in correlated electron systems is to find the new instability.  Namely, electrons need an attractive force to pair up, especially in both full and Fermi-arced pseudogap phases.

The attractive force to pair electrons is nothing but the \emph{quantum fluctuated} antiferromagnetic fluctuation.  Different from phonons, it is not the quantized entity of the antiferromagnetic fluctuation that is gapped of the energy $M_0$.  It is the quantum fluctuation of the amplitude $A_0$ given in Eq.~(\ref{e-field}).  The formation of the density modulation is the evidence of the attractive force to push electrons closer.  The formation of the Fermi arcs serves the evidence that the attractive force is larger than the repulsive force.

However, the attractive force is not in the direction of separation as the pairs stay in the nodal lines.  It is perpendicular to the separation.   Without the attractive force, the pairs break up and become the single-particle excitations.  Therefore, the form factor of the attractive potential in Fig.~(\ref{potential}b) indicates that the $d_{x^2-y^2}$, not the $d_{xy}$ or the $s$, symmetry is favored.  On the contrary, in the full pseudogap domains, electron pairs should have a full superconducting gap.

Another important ingredient of the electron pairs is the finite angular momentum.  To have it, a twist force must exist.  As the antiferromagnetic fluctuation is the longitudinal mode of the gauge field, there is also the transverse mode, that is spin-Berry's phase fluctuation~\cite{lee2018}.  The $B$ field of the spin-Berry's phase provides the twist force for the electron pairs in the way similar to the signal of the time-reversal-symmetry breaking fluctuation observed in the polar Kerr rotation experiments~\cite{lee2018}.  Without the quantum fluctuation of the $B$ field, $s$-wave paired electrons can be found in the full pseudogap domains but still not in the Fermi-arced domains.  With the quantum fluctuation of the $B$ field, $d_{x^2-y^2}\pm id_{xy}$-wave paired electrons can not be excluded from the theory in the full gap domains.  However, in the real situations, quantum fluctuations of the antiferromagnetic fluctuation and the spin-Berry's phase usually occur simultaneously.  The pairings of finite angular momentum are more likely than the $s$-wave.

Since the pairing is due to the attractive potential, the upper limit of the  superconducting gap can be estimated.  The effective attractive potential energy is the combination of the repulsive energy and the attractive energy, given by $\frac{\Delta}{4}-V_0$, where $\Delta$ is the pseudogap amplitude.  In the superconducting phase, the effective attractive potential energy is roughly the sum of the binding energy and the rotational kinetic energy.  The binding energy of the pair is $-2\Delta_s$, where $\Delta_s$ is the superconducting gap.  The rotational kinetic energy of a $d$-wave pair is roughly $2\frac{\hbar^2 l(l+1)}{2m^*r^2}$.  Taking $l=1$ and $r$ as one lattice constant, the rotational kinetic energy is 0.243 eV.  Estimating through the equation, $2\Delta_s+2\frac{\hbar^2 l(l+1)}{2m^*r^2} = V_0-\frac{\Delta}{4}$, the upper limit of the superconducting gap $\Delta_s$ is obtained as 55 meV.

Since the two kinds of the pseudogap domains can co-exist, the superconducting domains of different pairing symmetry can co-exist, as well, which is the main prediction of the theory.  The effective Hamiltonian can be given by 
\begin{eqnarray}
\mathscr{H}_j = -\frac{1}{2m^*}[\frac{\vec{\nabla}}{i}-g\vec{a}(x,y,t)]^2+V_j(x,y,t), \label{hami}
\end{eqnarray}
where $j=1, 2$ for the full and the Fermi-arced gap domains respectively, and $\vec{a}$ is the vector potential of the $B$ field fluctuation.  The interaction terms can be generated by using the standard perturbation techniques. For future directions, nonetheless, minimal models for different domains can be constructed in the lattice.

\section{Conclusion}
In general, insulators are believed to be stable against perturbations.  The superconductivity in the many-body insulators demonstrates the exceptional instability that seems to be as universal as the BCS instability in metals~\cite{sidorov2002, hosono2008}.  Quantum correlation, originating from the concept of wavefunction overlaps, plays the most important role.  Both the spatial wavefunction overlap (antiferromagnetic fluctuation) and the spin wavefunction overlap (spin Berry's phase fluctuation) dominate the electronic properties.  As the $d_{x^2-y^2}$ symmetry of the electron pairs reveals the most information in the mechanism of the superconductivity, the physics in the full pseudogap domains have been significantly ignored~\cite{yeh2001, feng2016}.  In particular, a new $d_{x^2-y^2}\pm id_{xy}$ symmetry of pairing is possible in those domains.  The new realm needs to be explored.

\section{Acknowledgement}
This work is funded by MOST 106-2112-M-002-007-MY3 of Taiwan.

\end{document}